\input packages.tx2	% all the packages together
% blackboard bold
\font\bbfont bbold10 at 12pt
\def\ZZ{\hbox{\bbfont Z}}
\def\QQ{\hbox{\bbfont Q}}
\def\RR{\hbox{\bbfont R}}
% Fraktur
\font\eufmtwelve eufm10 at 12pt
\def\GG{\hbox{\eufmtwelve G}}
\def\MM{\hbox{\eufmtwelve M}}
\def\LL{{\cal L}}
\def\Hom{\hbox{Hom}}
% definitions
\def\bra#1{{\langle#1|}}%		<#1|
\def\ket#1{{|#1\rangle}}%		|#1>
\def\bracket#1#2{{\bra{#1}#2\rangle}}%	<#1|#2>
\def\innerbracket#1#2{{\langle#1,#2\rangle}}%	<#1,#2>
\def\anglebrackets#1{\langle#1\rangle}% <#1>
\def\cochaind{\buildrel d\over{\buildrel\rightarrow\over{\vbox to.05em{}}}}
\def\chaind{\buildrel\partial\over{\buildrel\rightarrow\over{\vbox to.05em{}}}}

\twelvepoint
\baselineskip16pt

%%%%%%%%%%%%%%%%%%%%%%%%%%%%%%%%%%%%%%%%%%%%%%%%%%%%%%%%%%%%%%%%%%%%%%%%%%%%%%%%

\centerline{\bf Fourier-Space Crystallography as Group Cohomology}
\bigskip
\centerline{David A.\ Rabson}
\centerline{\it Department of Physics, PHY 114, University of South Florida,
Tampa, FL 33620, USA}
\smallskip
\centerline{Benji Fisher}
\centerline{\it Department of Mathematics, Boston College, Chestnut Hill,
MA 02467, USA}
\bigskip

We reformulate Fourier-space crystallography in the
language of cohomology of groups.  Once the problem is
understood as a classification of linear functions
on the lattice, restricted by a particular group relation, and identified
by gauge transformation, the cohomological description becomes natural.
We review Fourier-space crystallography and group cohomology, quote
the fact that cohomology is dual to homology, and exhibit several
results, previously established for special cases or by intricate
calculation, that fall immediately out of the formalism.  In particular,
we prove that {\it two phase functions are gauge equivalent if and only if
they agree on all their gauge-invariant integral linear combinations\/} and
show how to find all these linear combinations systematically.

\bigskip
\leftline{\tt PACS '01: 61.50.Ah, 61.44.Br, 61.44.Fw}

\vfil\eject

The discovery in 1984 of crystals with five-fold symmetry,\prlnote{%
D.\ Shechtman, I.\ Blech, D.\ Gratias, and J.W.\ Cahn,
Phys.\ Rev.\ Lett.\ \bf53\rm, 1951 (1984).}
and
therefore no periodicity, revived interest in
Bienenstock and Ewald's 1962 reformulation
of crystallography in Fourier space.\prlnote{%
A.\ Bienenstock and P.P.\ Ewald, Acta Cryst.\ {\bf15}, 1253 (1962).}
\ %
Mermin and collaborators have applied this Fourier-space crystallography
to the classification of the space groups of quasi-periodic
and periodic crystals,\prlnotez{%
D.S.\ Rokhsar, D.C.\ Wright, and N.D.\ Mermin, Phys.\ Rev.\ B {\bf37},
8145 (1988).}%
\ens{RWM}%
\prlnotez{%
D.A.\ Rabson, N.D.\ Mermin, D.S.\ Rokhsar, D.C.\ Wright,
Rev.\ Mod.\ Phys.~{\bf63}, 699 (1991).}%
\ens{RMP1}%
\prlnotez{%
N.D.\ Mermin,
Rev.\ Mod.\ Phys.~{\bf64}, 3 (1992);
Rev.\ Mod.\ Phys.~{\bf64}, 635[E] (1992);
Rev.\ Mod.\ Phys.~{\bf64}, 1163[E] (1992);
Rev.\ Mod.\ Phys.~{\bf66}, 249[E] (1944).}%
\ens{RMP2}%
\enrange{\enr{RWM}-\enr{RMP2}}
modulated crystals,\prlnote{%
N.D.\ Mermin and R.\ Lifshitz, Acta.\ Cryst.\ A {\bf48}, 515 (1992);
R.\ Lifshitz and N.D.\ Mermin, Acta.\ Cryst.\ A {\bf50}, 72 (1994);
Acta.\ Cryst.\ A {\bf50}, 85 (1994).}
and to color groups.\prlnote{%
R.\ Lifshitz, Rev.\ Mod.\ Phys.\ {\bf69}, 1181 (1997).}\ 
We show here that this technique, by now familiar to
crystallographers, has a particularly simple expression in the language of
homological algebra.

Section 1 quickly reviews the basis of Fourier-space
crystallography.  The {\it phase function\/}
$\Phi_g(\bold k)$, defined for $\bold k$ in the reciprocal
lattice and $g$ in the point group, facilitates the classification of
space groups.  We review the {\it gauge equivalence\/} of two phase
functions.  In section 2, we consider the necessary extinctions that
occur in the diffraction pattern when $g$ leaves $\bold k$ invariant, yet
$\Phi_g(\bold k)$ does not vanish.
Life would be simple if two phase functions
that agreed at all such pairs $(g,\bold k)$ were gauge-equivalent,
but in fact there are two crystallographic counterexamples;
however, when we consider gauge equivalence of {\it linear combinations\/} of
phase functions, we can state the result (Theorem 1) that two phase
functions are equivalent if and only if all their gauge-equivalent
linear combinations agree.  Section 3 reviews work by Mermin and K\"onig
on the most important such linear combination, one that
relates to a ray representation of the point group and
necessitates an
electronic degeneracy in the system.  Section 4 places the phase
function $\Phi$
in a {\it cohomology group}, while
section 5 puts gauge-invariant linear combinations of phase
functions in a related {\it homology group}.  In section 6, we invoke
the {\it duality\/} of these cohomology and homology groups to prove
Theorem 1 (an alternative, elementary, proof being presented in the appendix).
We also prove that for any phase function, there exists a gauge in which
it takes only rational values (with set denominator).
Section 7 applies our results to simplifying the classification of
space groups, and Section 8 addresses the ray representation of Section 3
in homological language.  Section 9 compares our results to a real-space
treatment.

\UH{1. Fourier-Space Crystallography}
Mermin and others have
argued persuasively\prlnotez{%
N.D.\ Mermin, Phys.\ Rev.\ Lett.\ {\bf68}, 1172 (1992).}%
\ens{Copernican}%
\prlnotez{%
N.D.\ Mermin, Phys.\ Stat.\ Sol.\ (a) {\bf151}, 275 (1995).}%
\ens{extinctions}%
\enrange{\enr{Copernican},\enr{extinctions}} that
the theoretical significance of quasicrystals lies
not so much in relaxing the requirement of periodicity
as in replacing exact {\it identity\/} of a density function
({\it e.g.}, electronic or nuclear)
under symmetry operations with {\it indistinguishability}.  We will
review Fourier-space crystallography only tersely, in order to establish
the notation; for a more developed exposition, we refer to \prlnotez{%
J.\ Dr\"ager and N.D.\ Mermin, Phys.\ Rev.\ Lett.\ {\bf76}, 1489 (1996).}%
\ens{Draeger-Mermin}%
Ref.\ [\enr{Draeger-Mermin}].

The {\bf reciprocal lattice}, $\LL$, (or simply {\bf lattice}, since
there is
no direct-space lattice of translations in the aperiodic cases) is the
closure under addition of the set of Bragg points $\bold k$
at which the Fourier transform $\rho(\bold k)$ of the direct-space density
function does not vanish.  Crystals, periodic or quasiperiodic, have
reciprocal lattices generated as all integral linear combinations of a finite
set of vectors $\bold k$; we do not consider non-crystals.
A different density function, $\rho'$, is
said to be {\bf indistinguishable} from $\rho$ if all spatially-averaged 
$n$-point correlation functions, representing macroscopic physical
measurements, are the same for the two densities: in Fourier space,
$$
\rho'(\bold k_1)\rho'(\bold k_2)\dots\rho'(\bold k_n)
=
\rho(\bold k_1)\rho(\bold k_2)\dots\rho(\bold k_n)
\eq(correl)
$$
for all lattice vectors summing to zero ($\sum\bold k_i=0$).
This holds if and only
if
$$
\rho'(\bold k)=e^{2\pi i\chi(\bold k)}\rho(\bold k)\smash{\quad,}
\eq(chi)
$$
where $\chi$, called a {\bf gauge function}, is any real-valued linear
function (taking values modulo 1)\prlnote{%
Henceforth, we shall understand the provision ``modulo unity'' to apply
to all such functions.  (Other workers in this field use the symbol $\equiv$
to denote equality modulo 1.)
}
on $\LL$. 
We are particularly interested
in the case that the first density function in \(chi)
results from acting on the second with an element $g$ of
a {\bf point group}, $G\subset\hbox{SO}(3)$, for then
$$
\rho'(\bold k) = \rho(g\bold k)
= e^{2\pi i\Phi_g(\bold k)}\rho(\bold k)\smash{\quad,}
\eq(phi)
$$
where $\Phi_g$, the particular gauge function for this $g$, is called
a {\bf phase function}.  Phase functions are constrained by the
{\bf group-compatibility} condition for $g,h\in G$,
$$
\Phi_{gh}(\bold k) = \Phi_g(h\bold k) + \Phi_h(\bold k)\smash{\quad,}
\eq(compat)
$$
a consequence of the associativity of the group action on the lattice.

The phase function defines algebraically the
action of the point group on the density and so, together with the lattice
and orientation, encodes the information
of the traditionally-defined {\bf space group}.
However, altogether too much freedom remains, for two phase functions,
$\Phi'$ and $\Phi$, related by a gauge function through
$$
\Phi'_g(\bold k) - \Phi_g(\bold k) = \chi(g\bold k-\bold k)
\eq(gauge)
$$
yield indistinguishable densities.  To classify space groups,
one therefore first computes all possible phase functions satisfying
\(compat), then groups into {\bf gauge-equivalence classes} phase
functions differing only by a gauge, as in \(gauge).\prlnote{%
To complete the classification,
one must also consider scale invariance and the orientation of $G$ with respect
to $\LL$. Ref.\ [\enr{Draeger-Mermin}] discusses in detail the classification of
{\bf Bravais classes}, {\bf arithmetic crystal classes}, and
{\bf space-group types}.  In
the present
work, we are concerned only with equivalence classes of phase functions.}
Until now, the second step has
required a clever choice of
specific gauge in which all but a few values of $\Phi_g(\bold k)$
vanish;
homological algebra provides a more systematic approach to the
classification, for it
it is precisely the group of functions satisfying \(compat) with the
identification \(gauge) that constitutes the fundamental object of
cohomology.

\UH{2. Extinctions, an untrue proposition, and a theorem}
The class containing the trivial $\Phi$, for which all values $\Phi_g(\bold k)$
vanish, is called {\bf symmorphic}.
Perhaps the most direct result of
Fourier-space crystallography concerns necessary extinctions in the
diffraction patterns of materials with certain non-symmorphic space groups.
The simplest example of the $n$-point correlation function in \(correl),
$\rho(\bold k)\rho(-\bold k)=|\rho(\bold k)|^2$, is proportional to the
intensity in a diffraction experiment.
As is well known, materials
with certain space groups show zero intensity at particular lattice
points $\bold k$.  From the Fourier-space point of view,
if some point-group element $g$ leaves $\bold k$ invariant, and if
$\Phi_g(\bold k)\not=0$,
\(phi) requires that
$\rho(\bold k)$ itself should vanish.  

By \(gauge), the condition
$g\bold k=\bold k$ means that all specimens $\Phi$ in a particular
equivalence class of phase functions will take the same value
for that choice of $g$ and $\bold k$, regardless of gauge.  We call the
particular value $\Phi_g(\bold k)$ a {\bf gauge-invariant part}.

By definition, if two phase functions are related by a gauge \(gauge), then
they agree on all
their gauge-invariant parts, but the converse is not true: out of the
230 periodic space groups in three dimensions, there
are precisely two non-symmorphic space groups with no non-zero
gauge-invariant $\Phi_g(\bold k)$.
Thus they agree with their symmorphic cousins
whenever $g\bold k=\bold k$,
yet they are not symmorphic.
Lacking non-zero gauge-invariants $\Phi_g(\bold k)$,
the two space groups, $I2_12_12_1$ and
$I2_13$, exhibit no systematic extinctions.
For the other 228 periodic space groups, and for all known quasiperiodic
space groups, it happens to be the case that {\it two phase functions
are related by a gauge if and only if they agree on all their gauge-invariant
parts of the form $\Phi_g(\bold k)$.}  If we knew this true in advance,
we could use it to simplify classification of equivalence classes of
phase functions by just finding all gauge-invariant parts compatible with
\(compat).
Absent this proposition, the gauge-invariant parts are insufficient: we
must find a clever choice of gauge in which all {\it non-}invariant 
phases vanish.

For the two peculiar cases in which one cannot make all non-invariant phases
vanish,
Mermin\enrange{\enr{RMP}} has constructed a gauge-invariant
{\it linear combination\/} of two phase functions,
$$
\Phi_g(\bold k_h) - \Phi_h(\bold k_g)
\eq(linearcombo)
$$
for a specific choice of commuting point-group elements $g, h$ and
lattice vectors $\bold k_h$ and $\bold k_g$, where neither $\Phi_g(\bold k_h)$
nor $\Phi_h(\bold k_g)$ is gauge-invariant.   By showing that the
group-compatibility condition \(compat) permits the gauge invariant
\(linearcombo) to assume a value either $0$ (symmorphic) or $1/2$
(non-symmorphic), he derives the existence of the two non-symmorphic 
space groups, $I2_12_12_1$ for point group $222$ on the orthorhombic $I$
lattice and $I2_13$ for point group $23$ on the cubic $I$ lattice.

The fact that
a gauge invariant of the form $\Phi_g(\bold k)$ is simply a linear
combination of one phase, while \(linearcombo) is a linear combination of two,
suggests a generalization of
the untrue converse about gauge-invariant single phases:

\penalty2000%HACK! prevent Theorem 1 from appearing at the top of a page
\smallskip%
{\narrower{{\bf Theorem 1.} Two phase functions are related by a gauge if and
only if they agree on all their gauge-invariant linear combinations, of
the form $\sum_i\Phi_{g_i}(\bold k_i)$.}\par}
\smallskip%

We will show (Section 5) that an appropriate homology group
systematically classifies all gauge invariants
and (Section 6) that its duality to a cohomology group proves the theorem.

\UH{3. Necessary degeneracies}
Although a non-zero value of the
invariant combination \(linearcombo) implies no necessary
extinctions in the diffraction pattern, K\"onig and Mermin\prlnote{%
A.\ K\"onig and N.D.\ Mermin, Phys.\ Rev.\ B {\bf56}, 13607 (1997);
Proc.\ Nat.\ Acad.\ Sci.\ USA {\bf96}, 3502 (1999);
Am.\ J.\ Phys. {\bf68}, 525 (2000).}
have given it a definite physical interpretation.
They consider a
reciprocal-space vector $\bold q$, not necessarily in $\LL$, and
the associated little group,
$G_{\bold q}\subseteq G$, consisting of elements $g$ such that
$$
\bold k_g\equiv\bold q-g\bold q
\eq(kg)
$$
{\it is\/} in the lattice.\prlnote{%
The special points $\bold q$ admitting non-trivial $G_{\bold q}$
fall into three categories: (1) lattice points,
(2) $\bold k_g\not=0$, which in the periodic case puts $\bold q$ on
a Brillouin-zone boundary, and (3) $\bold k_g=0$, {\it i.e.}, $\bold q$
invariant under $g$.  The first category
is uninteresting because \(linearcombo) always vanishes (for commuting
$g, h\in G_{\bold q}$)
as a consequence of \(compat).  The third category is similarly
uninteresting, since the corresponding phase must vanish.
}
The Hamiltonian $h_{\bold q}$
of the problem at Bloch wave-vector $\bold q$
describes all bands at $\bold q$; bands may repel or cross.
If for {\it commuting\/} $g$ and $h$ in $G_{\bold q}$, the
invariant \(linearcombo) does not vanish, every level must be degenerate with
some other level
at $\bold q$, so bands cross at least in pairs.  We
outline the argument, first because the gauge
invariant \(linearcombo) falls out of the first-homology calculation
below
and second because
it relates to a second cohomology group.

K\"onig and Mermin expand the Bloch wave in reciprocal-lattice vectors
$\bold k$ and express the
action of $g$ in the point group on $\bold k$ in terms of
the unitary operator $U(g)$, with
$$
U(g)\ket{\bold k} = e^{2\pi i\Phi_g(\bold k)}\ket{g\bold k-\bold k_g}
\smash{\quad.}
\eq(U)
$$
The operators $U(g)$ commute with the Hamiltonian $h_{\bold q}$ and
constitute a ray representation of $G_{\bold q}$, for
$$
U(g) U(h) = e^{-2\pi i\Phi_g(\bold k_h)}U(g h)\smash{\quad.}
\eq(ray)
$$
The map
$$
\lambda(g,h) = e^{-2\pi i\Phi_g(\bold k_h)}
\eq(factorsystem)
$$
from $G\times G$ to the unit circle is called the {\bf factor system}
at $\bold q$.

Suppose that $E$ is a non-degenerate energy level of $h_{\bold q}$, and let
$\ket{\psi}$ be the corresponding eigenvector.  Since the Hamiltonian
commutes with all the $U(g)$, $\ket{\psi}$ must also be an eigenvector
of each, so
$$
U(g)U(h) \ket\psi = U(h)U(g)\ket\psi\smash{\quad.}
\eq(commute)
$$
Comparing this commutation of matrices $U$ to
the ray representation \(ray), K\"onig and Mermin find,
for commuting $g$ and $h$,
that \(linearcombo) must vanish.  If it does not, every band must cross
at least one other band at $\bold q$.

\UH{4. Gauge-equivalence classes and the first cohomology group}
Brown\prlnote{%
K.S.\ Brown, {\it Cohomology of Groups}, Springer, New York, 1982.}\ens{Brown}
traces the theory of group cohomology to a 1904
work by Schur\prlnote{%
I.\ Schur, J.\ f\"ur die reine und angewandte Mathematik {\bf127}, 20 (1904);
reprinted in I.\ Schur, {\it Gesammelte Abhandlungen}, Springer, Berlin,
1973, v.\ 1, p.\ 86} on ray representations such as \(ray).  We
now use this theory to describe the classification of phase functions
up to gauge equivalence.  In Section 9, we compare this Fourier-space
cohomology to group cohomology in real-space crystallography.

The phase function $\Phi_g$, which maps the lattice $\LL$ to the
real numbers modulo unity, has three defining characteristics: (i) it is
linear, (ii) it satisfies the group-compatibility condition \(compat), and
(iii) it is defined only modulo gauge functions of the form \(gauge).
Cohomology conveniently packages functions of just such description.

The map $\Phi$, which takes $g\in G$ to $\Phi_g$, is an example
of a {\bf 1-cochain}, the set of all of which we call $C^1$.  We
denote by $\LL^*=\Hom(\LL,\RR/\ZZ)$ the group of all real-valued (mod 1) linear
functions on the lattice $\LL$;
the phase function $\Phi_g$ is in $\LL^*$.\prlnote{%
Conforming to mathematical notation, we use
$\Hom(A, B)$ to mean the set of linear functions
mapping $A$ into $B$,
$\hbox{\rm ker}(d)$ to indicate the kernel of a map $d$ ({\it i.e.}, the
set of arguments that $d$ maps to zero), $\hbox{\rm im}(d)$ for
its image ({\it i.e.}, if $d$ maps $A$ to $B$, that subset of $B$
that can be written as $d(a)$ for some $a\in A$),
\RR\ to denote the real numbers,
\ZZ\ the integers,
\QQ\ the rationals,
and $\RR/\ZZ$ the interval $[0,1)$.
}
More generally, an {\bf n-cochain} in $C^n$ is a function that
maps $n$ group elements from $G$ to $\LL^*$.
$C^n$ has a group structure under addition.
The gauge $\chi$, taking no point-group arguments, is a 0-cochain.

Connecting these functions is a {\bf coboundary} map,
$d:C^n\rightarrow C^{n+1}$,
such that
two maps in a row yield zero identically,
$d^{(n+1)}\circ d^{(n)}=0$.
Applied specifically to the lowest-order cases,%
\newbox\kludgebox\setbox\kludgebox\vbox{%
$$
\eqalign{
(d\sigma)(g_1, \ldots, g_{n+1}) ~=~& \sigma(g_1, \ldots, g_n)g_{n+1}
+ (-1)\sigma(g_1, g_2, \ldots, g_n g_{n+1})
+\ldots
\cr&+(-1)^n\sigma(g_1 g_2, g_3,\ldots,g_{n+1})
+(-1)^{n+1}\sigma(g_2, g_3, \ldots,g_{n+1})\quad.\cr%BROKEN:\smash{\quad.}\cr
}
$$
}%
\prlnote{
One convenient formulation of the general rule states
\smallskip\unvbox\kludgebox\vskip-0.5\baselineskip\noindent
By $\sigma(g)h = \sigma(g)\circ h$, one should
understand that group element $h$
acts on a member $\bold k$ of $\LL$ before the function $\sigma(g)$ then
acts on $h\bold k$.
}
\penalty-10000
$$
(d\chi)(g) = \chi\circ g - \chi
\eq(coboundary0)
$$
\nobreak
$$
\hbox{and}\quad (d\Phi)(g,h) = \Phi_g \circ h - \Phi_{g h}
+ \Phi_h\smash{\quad.}
\eq(coboundary1)
$$

Equation \(coboundary0) identifies those 1-cochains that are coboundaries
of 0-cochains as the group of
gauge transformations
(by \(gauge)); we define $B^n=\hbox{im}(C^{n-1}\cochaind C^n)$ as the
set of all {\bf n-coboundaries}.  Similarly, the right-hand side of
\(coboundary1) must vanish if $\Phi$ is to satisfy
the group-compatibility relation
\(compat); we define $Z^n=\hbox{ker}(C^n\cochaind C^{n+1})$ as the set
of all {\bf n-cocycles}, that is, n-cochains with vanishing coboundaries.

The quotient group $H^1=Z^1/B^1$, or cocycles modulo coboundaries,
is called the {\bf first
cohomology group} of $G$ with coefficients in $\LL^*$.
Membership in $Z^1$ establishes group-compatibility, while modding out
by $B^1$ identifies functions differing only by gauge transformation,
so $H^1$ {\it is exactly the group of equivalence classes of phase functions.}

\UH{5. Invariants and the first homology group}
To prove Theorem 1, we must find all gauge-invariant linear combinations
of phase functions.
Suppose the linear function $f:H^1\rightarrow\RR/\ZZ$ takes
an equivalence class of phase functions to a real number; we may think of
any gauge invariant in these terms.  Since the value of a gauge invariant
is the same for any member $\Phi$ of an equivalence class $\{\Phi\}$,
we may write $f(\{\Phi\})=f(\Phi)$ as only a slight abuse of notation.
We may turn this around to think of $\Phi$ mapping $f$ to a number.

To produce a number in the unit interval,
a cochain acts on integral linear combinations of ordered sets
each containing one lattice vector and $n$ group elements, which
we can write
(following Brown's convention\enrange{\enr{Brown}})
$\sum_i \bold k_i[g_{1i}|g_{2i}|\dots|g_{ni}]$ for $\bold k_i\in\LL$ and
$g_{ji}\in G$.  Since $\LL$ already
absorbs any integral coefficients, we need not write them explicitly.
Such linear combinations are called {\bf n-chains} and the set of
all $n$-chains $C_n=\LL[G^n]$.

A cochain $\Phi$ in $C^1$ and a chain
$c=\sum_i \bold k_i[g_i]$ in $C_1$ act on each other through the bracket
$$
\innerbracket{\Phi}{c}=\sum_i\Phi_{g_i}(\bold k_i)\in\RR/\ZZ
\smash{\quad.}
\eq(dot)
$$
We let $C_1$ inherit the additive
group structure of $\LL$:
$\bold k[g_1] + \bold k'[g_1] =
(\bold k+\bold k')[g_1]$.

By direct analogy with the coboundary operator in \(coboundary0)
and \(coboundary1), we define a {\bf boundary map},
$\partial:C_n\rightarrow C_{n-1}$, such that two maps in a row
yield zero.  Note that
$\partial$ decrements the number of copies
of $G$ in the chain, whereas the $d$ operator on cochains increments it.
In the lowest-order cases,%
\newbox\kludgekludgebox\setbox\kludgekludgebox\vbox{%
$$
\eqalign{
\partial(\bold k[g_1|g_2|\ldots|g_n]) =
&(g_n\bold k)[g_1|g_2|\ldots|g_{n-1}] + (-1)\bold k[g_1|g_2|\ldots|g_{n-1}g_n]
+ \ldots\cr &+(-1)^{n-1}\bold k[g_1 g_2|\ldots|g_n] +
(-1)^n\bold k[g_2|g_3|\ldots|g_n]\quad.\cr}%BROKEN:\smash{\quad.}\cr
$$
}%
\prlnote{%
The general recipe is
\smallskip\unvbox\kludgekludgebox\vskip-0.5\baselineskip\noindent
While this differs from Brown's formula, his duality theorem (7.4) can be
seen still to follow by reversing the order of the group elements $g_i$ and
replacing each with its inverse.
}\hfill\break
\penalty-10000
$$
\partial\bold k[g] = g\bold k - \bold k
\eq(boundary1)
$$%
\nobreak\vskip-2.2\baselineskip% hack away tex bug
$$%
\hbox{and}\quad\partial\bold k[g|h] =
(h\bold k)[g] - \bold k[gh] +
\bold k[h]\smash{\quad.}
\eq(boundary2)
$$

{\bf 1-cycles} are chains $c=\sum_i\bold k_i[g_i]$ satisfying
$$
\partial c = \sum_i ( g_i\bold k_i -\bold k_i) = 0\smash{\quad.}
\eq(cycle1)
$$
Let $Z_1=\hbox{ker}(C_1\chaind C_0)$ denote the set of all 1-cycles.
Applying the bracket \(dot) to $c$ satisfying \(cycle1),
we find that 1-cycles give
gauge-invariant linear combinations of phases, and since
the condition for gauge invariance of a {\it general\/} linear combination
of phases $\sum_i\Phi_{g_i}(\bold k_i)$ is, by \(gauge), precisely
\(cycle1), all such gauge-invariant linear combinations come from 1-cycles.
In the case that $c=\bold k[g]$ in \(cycle1),
the gauge invariant
takes the form $\Phi_g(\bold k)$ for
$g\bold k=\bold k$, but in general a linear combination, such as
\(linearcombo), must be allowed.

Similarly, {\bf1-boundaries} are those 1-chains that can be written as
boundaries of 2-chains, as in \(boundary2) or integral linear
combinations thereof; $B_1=\hbox{im}(C_2\chaind C_1)$.
Because $\partial\circ\partial=0$,
every 1-boundary is a 1-cycle, representing a gauge invariant.
Again applying the bracket
we discover that the phases at boundaries \(boundary2) are necessarily
zero by the group-compatibility condition \(compat).  Thus, although
1-boundaries are gauge invariants, they are trivial.
We mod them out in defining
the {\bf first homology group} of $G$ with coefficients in $\LL$:
$$
H_1 = Z_1 / B_1
\smash{\quad.}
\eq(homology)
$$
{\it First homology is
the group of non-trivial gauge invariants}.

We have already commented that a real-valued function $f$ on $H^1$
is unaffected by gauge ({\it i.e.}, by the addition of a coboundary to a
cocyle $\Phi$) and that
the addition of a boundary to a cycle $x$ does not change any phase.
These facts establish that the bracket, which we introduced in \(dot)
as a function $C^1\times C_1\rightarrow\RR/\ZZ$, is in fact
well defined on $H^1\times H_1$, which we may express concisely thus:
$$
\innerbracket{\Phi+d\chi}{\,x+\partial y} = \innerbracket{\Phi}{x}
\smash{\quad,}
\eq(dotH)
$$
where $\Phi$ is a phase function ($d\Phi=0$),
$d\chi$ a gauge transformation of the
form \(gauge), $x\in C_1$ a cycle ($\partial x=0$), and $\partial y$ a boundary,
$y\in C_2$.

\UH{6. Duality and the Proof of Theorem 1}
We note a striking similarity between $H^1$ and $H_1$.  While the former
contains all linear functions satisfying group compatibility, identifying those
related by gauge
transformation, the latter contains all linear gauge invariants, identifying
any whose
difference is
made trivial by group compatibility.  In fact, homological algebra
affirms their duality.  Before stating this central result,
we note the following:

{\narrower{{\bf Lemma 2.} For $G$ a finite group with $|G|$ elements and
$\Phi\in H^1$, $|G|\Phi=0$ (where $0\in H^1$ denotes the set of phase functions
gauge-equivalent to zero).  In other words,
for any phase function $\Phi$, there exists a
gauge in which every
$\Phi_g(\bold k)$
is rational with denominator $|G|$.
}\par}

This is quite plausible, since the only constraints on phases are those
imposed by group compatibility \(compat) applied through the constituent
relations of the point group.  Specifically, summing over $g$ in \(compat)
and noting that $\sum_g\Phi_{gh}=\sum_g\Phi_g$, we find
$$
|G|\Phi_h(\bold k) = \chi(h\bold k -\bold k)\smash{\quad,}
\eq(lemma2)
$$
where $\chi(\bold k)=-\sum_g\Phi_g(\bold k)$ clearly satisfies the
definition of a gauge function.  Then $|G|$ times any phase function
is a pure gauge, proving the lemma.  Note that we may {\it not\/} divide
by $|G|$ to show that all phase functions are gauge-equivalent to zero,
because
$\chi$ is defined only modulo unity.

Having found a gauge in which a phase function takes rational values, we define
$\LL'=\Hom(\LL,\QQ/\ZZ)\subseteq\LL^*$ and quote the following standard
result from cohomology:\prlnote{Brown, {\it op.\ cit.}, theorem 7.4, p.\ 147.
A ``free Abelian'' group (represented additively)
is one for which no non-trivial integral
linear combination of a minimal set of generators
vanishes.}

{\narrower{{\bf Lemma 3.} For $G$ a finite group and $\LL$ a
free Abelian group, the groups $H^n(G,\LL')$ and
$H_n(G,\LL)$ are dual through
the bracket \(dot)
$\innerbracket{\,}{}: H^n \times H_n
\rightarrow \QQ/\ZZ$.
}\par}

Duality means that if $X$ serves as a minimal basis for $H_1$, there
exists a basis $\{\Phi^x\}$ for $H^1$ labeled by $x\in X$ and that
any $\Phi\in H^1$ can be decomposed uniquely as a linear combination of
the basis elements $\Phi^x$ with
integral coefficients equal to
$\innerbracket{\Phi}{x}/\innerbracket{\Phi^x}{x}$.
Therefore, if we know the value
$\innerbracket{\Phi}{x}$ of a phase function on each of the non-trivial
gauge invariants $x$ spanning $H_1$, by linearity we know $\Phi$ up to
gauge transformation.  This proves Theorem 1.

We give in the appendix an alternative, elementary proof
of Theorem 1
by constructing
the gauge function relating two phase functions that agree on all
gauge-invariant linear combinations.

\UH{7. Calculating Space Groups (example)}
Homological algebra offers a powerful tool for systematically finding
all gauge invariants.  By Theorem 1, it furthermore frees us from having
to specify a gauge: once we have the gauge invariants, we no longer
have to prove on a case-by-case basis that all other phases can be made
to vanish.  As an example, we compute the two possible space-group classes
on the body-centered orthorhombic direct-space lattice (which is
face-centered in reciprocal space) with the point group $222$.  Instead of
having to hunt for the gauge invariant \(linearcombo), we shall see it fall
out of the formalism naturally.

The calculation is straightforward.  First we calculate
the group $Z_1$ of gauge invariants by acting with the boundary operator
$\partial$ on $1$-chains $\bold k[g]$.  We then wish to remove from the list all
boundaries in $B_1$, since these represent the ``trivial'' consequence
of group compatibility \(compat); to do so, we let $\partial$ act on
$2$-chains $\bold k[g|h]$.  After this elimination, we find a homology group
with only two elements.
The entire procedure is easily
automated; for example,
the {\tt NullSpace[]} and {\tt LatticeReduce[]} functions of Wolfram's
Mathematica program can be used to find minimal bases over $\ZZ$ for
$Z_1$ and $B_1$.

Let ${\bf\hat e}^i$, $i=1,2,3$, constitute the usual Cartesian axes.
We name the three sides of the conventional reciprocal cell
$\bold x=a{\bf\hat e}^1$, $\bold y=b{\bf\hat e}^2$, and
$\bold z=c{\bf\hat e}^3$, where
$a$, $b$, and $c$ are all different,
and generate the face-centered
reciprocal lattice $\LL$ (dual to the body-centered direct
lattice) with the vectors $b_1={1\over2}(011)$, $b_2={1\over2}(101)$, and
$b_3={1\over2}(110)$ (all in terms of the $(\bold x,
\bold y, \bold z)$ basis).
The point group is $G=\{e,d_1,d_2,d_3\}$, where $d_j$ leaves ${\bf\hat e}^j$ 
invariant, $d_1 d_2=d_3$ {\it et cycl.}, and $d_i^2=e$, the identity.
$G$ is generated by $d_1$ and $d_2$.

To calculate the group of gauge invariants, $Z_1$,
we apply $\partial$
to the (additive) generators of 1-chains.  The chains $\bold k[e]$ are
cycles, since $e$ leaves all vectors invariant.  We can show that they are
also boundaries, for
by group compatibility,
$$
\bold k[gh] \doteq (h\bold k)[g] + \bold k[h]\smash{\quad,}
\eq(kgh)
$$
where $\doteq$ denotes equality up to the boundary $\partial\bold k[g|h]$.
Setting $g=h=e$, we then have $\bold k[e]\doteq0$, so we leave it out of the
computation.  Equation \(kgh) furthermore tells us that
we need consider only the
generating elements $d_1$ and $d_2$, not $d_3=d_1 d_2$.

One easily verifies that
$$
(\partial\bold b_i[d_j])_k = |\epsilon_{ijk}| - \delta_{ik} - \delta_{jk}
\smash{\quad,}
\eq(p3tensor)
$$
where the subscript $k$ indicates the resulting lattice component in
the $(\bold b_1, \bold b_2, \bold b_3)$ basis and where $\delta$
is the Kronecker delta and $\epsilon$ the totally antisymmetric tensor.

With $i=1,2,3$ and $j$ restricted to $1,2$, we can think of the left-hand
side of \(p3tensor) as a three-by-six matrix whose three-dimensional null
space over the integers is $Z_1$ (aside from the boundaries we have
already eliminated).  It is a standard result of linear algebra that
row reduction without division yields a primitive basis over the
integers;\prlnote{%
N.\ Jacobson, {\it Basic Algebra I}, Freeman, San Francisco, 1974;
see section 3.7.}
we find for this basis
$$
\eqalign{
z_1 &= \bold x[d_1]\cr
z_2 &= \bold y[d_2]\cr
z_3 &= -\bold b_1[d_2] + \bold b_2[d_1]\smash{\quad.}\cr
}
\eq(6z1)
$$
We now know all the gauge invariants and seek, by computing $B_1$, to identify
those that are related by group compatibility \(compat).

To find the
1-boundaries not already identified, we must
apply $\partial$ to the 27 two-chains
$$
\bold b_i[d_j|d_k]\smash{\quad;}
\eq(2chain)
$$
this number is reduced to 18 if we notice that
$\bold k[g|h_1 h_2]$ has the same boun\-da\-ry as
$(h_2\bold k)[g|h_1] \penalty-5000+
\bold k[g h_1|h_2] - \bold k[h_1|h_2]$, so that
in \(2chain) we may restrict $k$ to $1, 2$.

After eliminating duplications and zeroes, we get seven boundaries.
Rearranging the rows into echelon form, we find only three that
are linearly independent: $z_1$, $z_2$, and $2z_3$ are
all boundaries.
This identifies all {\it even\/} multiples of the cycle
$z_3$ in $H_1$, which therefore contains just a single non-zero invariant:
$$
H_1(G,\LL)=\{0,z_3\}\smash{\quad.}
\eq(212121)
$$
Since $2z_3\doteq0$, $\Phi(z_3)$ can take values only $0$ and $1/2$.

The invariant $z_3$ is precisely the linear combination \(linearcombo)
of phases used in Ref.\ [\enr{RMP2}] to
distinguish the symmorphic ($I222$, $\Phi(z_3)=0$)
from the non-symmorphic ($I2_12_12_1$, $\Phi(z_3)=1/2$) space group.
Mermin's rather shorter calculation rests on a clever choice of gauge
ensuring
$\Phi_{d_1}(\bold b_3) = \Phi_{d_2}(\bold b_1) = \Phi_{d_3}(\bold b_2) = 0$;
in our comparatively pedestrian approach, the list of phases that can be
discarded inheres in the calculation of the boundaries $B_1$.  Significantly,
the present calculation is easy to automate, requiring no ``judicious choice''
of generators and gauges, while verification that one gets ``a representative of
{\it every\/} class''\enrange{\enr{RMP2}} is built in to the procedure.
We lose the elegance of Mermin's (equivalent)
calculation, but we hope we make up for
some of the loss in the compact and general statement of Theorem 1 as
proven in Section 6.  We have generalized and applied essentially the
ideas demonstrated in this section to the simultaneous computation of a large
number of cases, including non-standard and modulated
lattices.\prlnote{%
B.\ Fisher and D.A.\ Rabson, {\it Applications of Group Cohomology to
the Classification of Crystals and
Quasicrystals}, unpublished, {\tt math-ph/0105010}.}

\UH{8. Second Cohomology and the Factor System}
In general, the cohomology group $H^n({\GG,\MM})$ consists of cocycles from
$n$ copies of a group $\GG$ to a module $\MM$ with the identification of
coboundaries.  Having classified the phase functions in
$H^1$ with $\GG=G$ and $\MM=\LL'$, we can
calculate the factor system \(factorsystem).  However, the factor system
also has a cohomological existence in its own right, with a different
group $\GG$ and different module $\MM$.  Consider
$$
\Lambda(g,h)=-(\log\lambda)/(2\pi i)=\Phi_g(\bold k_h)\smash{\quad,}
\eq(Lambda)
$$
taking values in the module $\MM=[0,1)=\RR/\ZZ$.  From here on, we will use the
term ``factor system'' to refer to $\Lambda$ rather than to $\lambda$.
Since $g$ and $h$ live in the little group $\GG=G_{\bold q}$, $\Lambda$ is
a 2-cochain in $C^2(G_{\bold q},\RR/\ZZ)$.  The action of $g\in G_{\bold q}$ on
$x$ in the module $\RR/\ZZ$ is trivial: $xg=x$.

We can again impose group compatibility as a cocycle condition:
$$
(d\Lambda)(g_1,g_2,g_3)\equiv\Lambda(g_1,g_2)-\Lambda(g_1,g_2g_3)
+\Lambda(g_1g_2,g_3)-\Lambda(g_2,g_3)=0\smash{\quad.}
\eq(cocycle2)
$$
(The equivalence to
group compatibility follows from the definitions \(kg) and \(Lambda).)\prlnote{%
The astute will notice the absence
of a $g_{n+1}$ element acting to the left on the first term.  This
is because of the trivial group action
$\Lambda(g_1,g_2)\circ g_3=\Lambda(g_1,g_2)$.
}

If two ray representations \(U) differ
by a quantum phase $e^{2\pi i\sigma(g)}$
(one for each $g\in G_{\bold q}$) so that $U'(g)=e^{2\pi i\sigma(g)}U(g)$,
the corresponding factor systems
differ by the coboundary of $\sigma$: 
$$
\Lambda(g,h) - \Lambda'(g,h) = (\delta\sigma)(g,h) \equiv
\sigma(g) - \sigma(gh) + \sigma(h)\smash{\quad.}
\eq(Lambdacoboundary)
$$
Since the quantum phase
has no physical consequence, we identify ray representations
related by \(Lambdacoboundary).  {\it The set of phase-equivalence classes
of ray representations is the\/ {\bf second cohomology group},
$H^2(G_{\bold q},\RR/\ZZ) = Z^2/B^2$.}

As a final application of group cohomology, we consider a lattice
vector $\bold k$ left invariant by a point-group element $g\in G$.
Letting $\GG=\anglebrackets{g}=\{g^n|n\in\ZZ\}$, the
cyclic group generated by $g$,
and $\MM=\LL$, we express $\bold k\in\LL$ as a 0-cochain.  The cocycle
condition $(d\bold k)(g^n)\equiv g^n\bold k -\bold k=0$
expresses the requirement
that $\bold k$ be fixed by $g$.  Since there are no ($-1$)-cochains, there
is no identification of coboundaries, and the cohomology $H^0=Z^0$ is
the group of all $\bold k$ left invariant by $g$.

\UH{9. Comparison to Real-Space Cohomology}
It has long been known that cohomology groups can be used to classify the
space groups of periodic crystals; Hiller\prlnote{%
H.\ Hiller, Amer.\ Math.\ Monthly {\bf93}, 765 (1986).}%
\ens{Hiller}
gives a particularly approachable exposition.
We now
compare our formulation, using the
Fourier-space lattice $\LL$, to the traditional
approach, described in terms of the periodic lattice~$L$ in real space.
In terms of~$L$, we can describe~$\LL$ as the lattice of integer-valued
homomorphisms on~$L$,
$ \LL = \Hom(L,\ZZ) $.
(For mathematical convenience, and to allow direct comparison with
Hiller, we have in this section absorbed the conventional factor
of $2\pi$ into $\LL$.)
In the three-dimensional case,
if we choose a basis
$\bold a_1$,~$\bold a_2$, $\bold a_3$
of~$L$ then we can describe~$\LL$ in terms of the dual basis
$\bold k_1$,~$\bold k_2$, $\bold k_3$,
where
$ \bracket{\bold k_i}{\bold a_j} = \delta_{ij} $
(Kronecker delta).

Any real-valued linear form on~$\LL$ has the form~$\bracket{\bold k}{\bold x}$
for a fixed
vector~$\bold x$ in real space.  Recall that our phase functions~$\Phi_g$ take
values in~$\RR/\ZZ$, not~$\RR$:  that is, we identify~$\Phi_g(\bold k)$
with~$\Phi_g(\bold k)+n$, for
any integer~$n$.  This corresponds to identifying~$\bold x$
with~$\bold x+\bold a$, for any
$\bold a \in L$.
Thus there is a one-to-one correspondence between phase functions and elements
of the quotient group~$\RR^3/L$, and we can write
$ \Phi_g(\bold k) = \bracket{\bold k}{\bold x_g} $
for some real-space vector~$\bold x_g$, identified with any of its translates by
lattice vectors.

In Fourier-space crystallography, the phase functions~$\Phi_g$ are required to
satisfy the group-compatibility condition~\(compat), or equivalently the cocycle
condition
$ (d \Phi)(g,h) = 0 $
({\it cf.}~\(coboundary1)).
In terms of the vectors~$\bold x_g$ in real space, this condition
becomes
$ \bracket{\bold k}{\bold x_{gh}} = \bracket{h\bold k}{\bold x_g} +
\bracket{\bold k}{\bold x_h} $.
Recall what it means to let~$h$ (an orthogonal transformation on real space)
act on~$\bold k$ (a vector in Fourier space):  by definition,
$ \bracket{h\bold k}{\bold x} = \bracket{\bold k}{h^{-1} \bold x} $.
Thus the group-compatibility condition becomes
$$
  \bold x_{gh} = h^{-1} \bold x_g + \bold x_h \smash{\quad.}
\eq(Hillerx)
$$
In order to compare this with the notation of Ref.\ [\enr{Hiller}], let
$ s(g) = \bold x_{g^{-1}} $
and replace
$g$~and~$h$
with
$g^{-1}$~and~$h^{-1}$,
respectively:
$$
  s(hg) = \bold x_{g^{-1} h^{-1}} = \bold x_{h^{-1}} +
h \bold x_{g^{-1}} = s(h) + h \, s(g)
  \smash{\quad.}
\eq(Hillery)
$$
This is exactly the cocycle condition used in Ref.\ [\enr{Hiller}], Prop.~5.1.

Finally, consider gauge equivalence: we identify phase functions
$\Phi_g$~and~$\Phi'_g$
related by a gauge transformation as in~\(gauge).  In the periodic case we are
considering, the gauge function can be described by
$ \chi(\bold k) = \bracket{\bold k}{\bold y} $,
where again $\bold y$~is a
real-space vector considered modulo~$L$.  If the second
phase function~$\Phi'_g$ corresponds to~$\bold x'_g$ in real space then the
condition~\(gauge) can now be expressed as
$ \bracket{\bold k}{\bold x'_{g}}
- \bracket{\bold k}{\bold x_{g}} = \bracket{g\bold k-\bold k}{\bold y} $,
or
$ \bold x'_g - \bold x_g = g^{-1} \bold y - \bold y $.
In terms of
$ s(g) = \bold x_{g^{-1}} $
and
$ s'(g) = \bold x'_{g^{-1}} $,
this becomes
$$
  s(g) - s'(g) = \bold y - g\bold y \smash{\quad,}
\eq(Hillerz)
$$
which is exactly the coboundary condition in Ref.~[\enr{Hiller}].

In conclusion, the cohomology group
$ H^1 = Z^1 / B^1 $
that we consider here agrees, in the periodic case, with the cohomology group
$ H^1(G, \RR^3/L) $
described in Ref.~[\enr{Hiller}].

\UH{Acknowledgements}
D.R.\ gratefully acknowledges the hospitality of the mathematics department
at Boston College, where some of this work was done.

\UH{Appendix: Elementary Proof of Theorem 1}
We have already shown that duality proves Theorem 1, which we may restate,
{\it a phase function is gauge-equivalent to zero if and only if it
vanishes on all gauge-invariant linear combinations}, or again
equivalently, {\it a 1-cocycle is a coboundary iff it vanishes on all cycles}.
The implication (``only if'') statements come immediately, so we
prove the converse.

Although standard treatises on homological algebra prove duality of $H^n$
and $H_n$ (Lemma 3), the following elementary demonstration
of Theorem 1 also explicitly constructs the gauge transformation $\chi$
for which $\Phi_g(\bold k)=\chi(g\bold k-\bold k)$ if
$\Phi$ vanishes on all gauge invariants.
The proof extends familiar ideas
from the linear algebra of vector spaces to that of free modules
(over $\ZZ$).%  We use the result that such modules have bases.

Define the linear function $\tilde\chi$ on all
0-boundaries
$\partial c\in B_0$ ($c\in C_1$) by
$$
\innerbracket{\tilde\chi}{\partial c}\equiv\innerbracket{\Phi}{c}
\smash{\quad.}
\eq(chitilde)
$$
We can do so consistent with linearity because, by hypothesis,
$\innerbracket{\Phi}{z}=0$ for all gauge invariants $z\in\hbox{ker}\,\partial$.
As a special case of \(chitilde), for $c={\bold k}[g]$, we have
$\Phi_g(\bold k)=\tilde\chi(g\bold k-\bold k)$, which almost makes $\Phi$
gauge-equivalent to zero.
However, to establish
that $\Phi$ is purely a gauge (thus proving the theorem), we must extend
$\tilde\chi$, which we have so far defined only on $B_0$, to
a linear function $\chi$ defined on all $\LL=C_0$.
To extend $\tilde\chi$ to $\chi$ on $C_0$, write the rectangular
matrix whose $n_c=\hbox{rank}(B_0)$ columns are the integral expansions of
a minimal basis of $B_0$ in a basis of $C_0$; $M$
has $n_r=\hbox{rank}(C_0)\geq n_c$ rows.  $M$ maps $B_0$ to $C_0$ by inclusion.

$M^TM$ is a $n_c\times n_c$ matrix; if $M^TM$ is not singular, we can
form the left inverse of $M$ as $\tilde M=(M^T M)^{-1} M^T$.  To show
$M^TM$ non-singular, observe that the columns of $M$ are all independent over
the integers,
so that we can find a matrix $M'$ with $n_r$ rows and $n_r-n_c$ columns,
all columns orthogonal to those of $M$.\prlnote{%
Since the columns of $M$ are independent over $\ZZ$, they are independent
over $\QQ$, which is a field, so we can choose the columns of $M'$ not just
independent of those of $M$ but also orthogonal.}
The $n_r\times n_r$ square matrix
$A=(M M')$ is therefore non-singular.  Since $A$ and $A^T$ are both
non-singular, so is $A^TA=\left(\matrix{M^TM&0\cr0&M'^TM'\cr}\right)$,
but since $\hbox{det}(A^TA)\not=0$, we must have $\hbox{det}(M^TM)\not=0$.

Think of $\tilde\chi$ as a row vector of rank $n_c$; define
$\chi=\tilde{\vphantom{M}\chi}\tilde M$ (row
vector of rank $n_r$).  This completes
the construction of a linear $\chi$ on all $C_0$ such that
$\Phi_g(\bold k)=\chi(g\bold k-\bold k)$, proving Theorem 1.

\UH{References}
\endnotes

\end